\documentclass[12pt,english]{article}
\usepackage[T1]{fontenc}
\usepackage[latin9]{inputenc}
\usepackage{setspace}
\usepackage{amssymb}

\makeatletter
\newcommand{\lyxaddress}[1]{
\par {\raggedright #1
\vspace{1.4em}
\noindent\par}
}

\makeatother

\usepackage{babel}

\begin{document}

\title{Nuclear and Particle Physics applications of the Bohm Picture of
Quantum Mechanics}

\author{A. Miranda}

\maketitle

\lyxaddress{Department of Physics and Astronomy, Aarhus University, DK-8000 Aarhus
C, Denmark}

\begin{abstract}
Aproximation methods for calculating individual particle/ field motions
in spacetime at the quantum level of accuracy (a key feature of the
Bohm Picture of Quantum Mechanics (BP)), are studied.

This sharply illuminates not only the deep quantum structures underlying
any observable quantum statistical laws of motion of particles and
fields in spacetime, but also how the continuous merging of the so-called
classical and quantal modes of description actually occurs, with no
breaks anywhere.

Modern textbook presentations of Quantum Theory are used throughout,
but only to provide the necessary, already existing, tested formalisms
and calculational techniques. New coherent insights, reinterpretations
of old solutions and results, and new (in principle testable) quantitative
and qualitative predictions, can be obtained on the basis of the BP
that complete the standard type of postdictions and predictions.

Most of the dead wood still cluttering discussions on the meaning
of Quantum Theory and the role of the BP is by-passed.

We shall try to draw attention to the physics of this unfortunately
hardly known novel formulation of Quantum Theory by giving additional
illustrative examples inspired from the daily practices of contemporary
Nuclear and Particle Physics, subjects that as yet have not been thoroughly
reinterpreted within the BP.

These fields of research offer excellent oppurtunities for explaining
and illustrating the significance of time in quantum transitions,
as well as the closely related features of quantum non-locality and
quantum wholeness, as hard physical facts. We claim that in addition
we can obtain a substantial gain in predictive powers of the underlying,
all-encompassing, Quantum Theory.

\end{abstract}

\section{A short introduction to the Bohm Picture of Quantum Mechanics (BP)}

Quantum Theory, as it is explained in textbooks, comes in two fundamental
versions, classified by Dirac with the word \emph{Pictures} {[}1]:
the Schroedinger Picture (SP), where time developments are entirely
accounted for by the wavefunctions (which can be described in many
different \emph{representations}) and the Heisenberg Picture (HP),
where time-developments are entirely encoded in the operators.

\subsection{Starting from the SP}

The differential form of the Schroedinger equation of motion is (we
set $\hbar=1,$ unless otherwise stated):

\begin{equation}
i\frac{d}{dt}|\Psi(t)>=H|\Psi(t)>\label{eq:(1)}\end{equation}

An integral representation of the Schroedinger ket vector $|\Psi(t)>$
is

\begin{equation}
|\Psi(t)>=e^{-iH(t-t')}|\Psi(t')>\label{eq:(2)}\end{equation}

The Hamiltonian H in this context has a double role: it is the operator
generating time-translations, assumed to be itself time-independent
(for isolated systems), and it is the operator whose eigenvectors
are stationary states $\Phi_{\nu}$ of the system:

\begin{equation}
H(N)\Phi_{\nu}(\{\vec{x}\};E_{\nu})=E_{\nu}\Phi_{\nu}(\{\vec{x}\};E_{\nu})\label{eq:(3)}\end{equation}

We shall use the simplifying short-hand notation:

\begin{equation}
\{\vec{x}\}\equiv\vec{x}_{1},\vec{x}_{2},...,\vec{x}_{N}\label{eq:(4)}\end{equation}

\begin{equation}
\{\vec{p}\}\equiv-i\vec{\nabla}_{1},-i\vec{\nabla}_{2},...,-i\vec{\nabla}_{N}\label{eq:(5)}\end{equation}

In general, let us assume some complete set of state vectors in Hilbert
space:

\begin{equation}
\sum_{\alpha}|\alpha><\alpha|=1\label{eq:(6)}\end{equation}

\begin{equation}
<\beta|\alpha>=\delta(\beta,\alpha)\label{eq:(7)}\end{equation}

Then the wavefunctions (or their complex conjugates) are just components
(complex numbers) of the ket vector $|\Psi(t)>($or the bra vector
$<\Psi(t)|$) in the Hilbert space in whatever the chosen basis vectors
is {[}1] :

\begin{equation}
\Psi_{\beta}(t)=<\beta|\Psi(t)>\label{eq:(8)}\end{equation}

\begin{equation}
\Psi_{\alpha}^{*}(t)=<\Psi(t)|\alpha>\label{eq:(9)}\end{equation}

We work in this paper with an entirely different Picture, unfortunately
not even mentioned in textbooks. To give it a name, we shall refer
to it as the Bohm Picture (BP){[}2,3,4] and shall be discussed presently.

Before we proceed, let us ask the question : would it make any real
difference, had Bohm started instead from the HP, instead of the SP? 

From one of Dirac's last publications {[}1] we can infer that for
non-relativistic systems the answer is NO; for relativistic Quantum
Field Theory (QFT), the case seems to remain open. 

Practical approximation methods for solving the time-dependent and
time-independent Schroedinger equation, for reallistic systems, both
in the textbook formulation and in published BP papers {[}2,3], are
generally based on some standard combinations of Perturbation - Variation
theory with computer power. 

As we shall see, wavefunctions associated with specific physical systems
are an essential part of input information, but only a part. This
statement sharply disagrees not only with textbook quantum theory
but also with all (as far as I know) alternative ontological formulations
{[}2].

Let us attempt to look more closely into the question of time development
in Quantum Mechanics, as accounted for in the BP. As it is perhaps
widely agreed, an universal time seems to be required by the very
definition of the Schroedinger equation.

We follow Bohm by choosing the basis set of wavefunctions to be the
eigenstates of the Schroedinger (hermitean) {}``position operators''
$\hat{x}_{k}$: \begin{equation}
\hat{x}_{k}|\vec{x}_{1},\vec{x}_{2},...,\vec{x}_{N}>=\vec{x}_{k}|\vec{x}_{1},\vec{x}_{2},...,\vec{x}_{N}>\label{eq:(10)}\end{equation}

\begin{equation}
[\hat{x}_{k},\hat{x}_{j}]=0\label{eq:(11)}\end{equation}

\begin{equation}
[\hat{p_{k}},\hat{p_{j}}]=0\label{eq:(12)}\end{equation}

\begin{equation}
[\hat{x}_{k},\hat{p_{j}}]=i\delta(k,j)\label{eq:(13)}\end{equation}

\begin{equation}
\Psi(\{\vec{x}\};t)\equiv<\{\vec{x}\}|\Psi(t)>\label{eq:(14)}\end{equation}

The time-dependent Schroedinger equation for describing a system with
countable N=1,2,..., degrees of freedom can always be rewritten in
the form\begin{equation}
H(N)\Psi(\{\vec{x}\};t)=i\frac{\partial}{\partial t}\Psi(\{\vec{x}\};t)\label{eq:(15)}\end{equation}
where H(N) is the (supposed hermitean) Hamiltonian:\begin{equation}
H(N)=U_{0}-\sum_{i=1}^{N}\frac{1}{2m_{i}}\nabla_{i}^{2}+V(\{\vec{x}\})\label{eq:(16)}\end{equation}

$U_{0}$ is an arbitrary constant. The function V represents a $preassigned$
classically definable $local$ function:

\begin{equation}
<\{\vec{x}'\}|V|\{\vec{x}\}>\equiv\delta(\{\vec{x}'\}-\{\vec{x}\})V(\{\vec{x}\})\label{eq:(17)}\end{equation}

So a general time-dependent wavefunction is obviously \begin{equation}
\Psi(\{\vec{x}\};t)=\sum_{\nu,E_{\nu}}C_{\nu}(E_{\nu})\Phi(\{\vec{x}\};E_{\nu})\exp(-iE_{\nu}t)\label{eq:(18)}\end{equation}
and is completely defined, if the values of the coefficients $C_{\nu}$
be given at some initial time t=0:\begin{equation}
\Psi(\{\vec{x}\};0)=\sum_{\nu,E_{\nu}}C_{\nu}(E_{\nu})\Phi(\{\vec{x});E_{\nu})\label{eq:(19)}\end{equation}
Thus assuming that an initial wave-packet is given, we immediately
deduce an explicit expression for the coefficients\begin{equation}
\Psi(\{\vec{x}\};0)=\sum_{i}a_{i}\varphi_{i}(\{\vec{x}\})\label{eq:(20)}\end{equation}
\begin{equation}
\Phi_{\nu}(\{\vec{x}\},E_{\nu})=\sum_{i}b_{\nu i}(E_{\nu})\varphi_{i}(\{\vec{x}\})\label{eq:(21)}\end{equation}
\begin{equation}
C_{\nu}(E_{\nu})=\sum_{i}b_{\nu i}^{*}(E_{\nu})a_{i}\label{eq:(22)}\end{equation}
where $\Psi$and $\Phi$ are expanded over some convenient complete
set of functions $\left\{ \varphi_{i}\right\} $. 

We shall simply assume familiarity with this standard material and
carry on from there. Let us next introduce the BP on the basis of
the following postulates:

POSTULATE 1

As the wavefunction is in general a complex number, let us write\begin{equation}
\Psi(\{\vec{x}\};t)=R(\{\vec{x}\};t)e^{iS(\{\vec{x}\};t)}\label{eq:(23)}\end{equation}
where R and S are both real.

This by definition a $beable$ {[}2,3,4] oficially known {}``wavefunction''
. It obeys its own equation of motion, which is none other than the
Schroedinger equation. The latter can be rewritten as a system of
time-dependent coupled differential equations involving two real functions
R and S:

\[
\frac{\partial S(\{\vec{x}\};t)}{\partial t}+\]

\begin{equation}
+\sum_{k=1}^{N}\frac{1}{2m_{k}}|\vec{\nabla}_{k}S(\{\vec{x}\};t)|^{2}+V(\{\vec{x}\})+Q(\{\vec{x}\};t)\label{eq:(24)}\end{equation}
\begin{equation}
\frac{\partial P(\{\vec{x}\};t)}{\partial t}+\vec{\nabla}.{\displaystyle {\displaystyle \sum_{k}}}\vec{J}_{k}(\{\vec{x}\};t)=0\label{eq:(25)}\end{equation}

with the definitions

\begin{equation}
\vec{J}_{k}(\{\vec{x}\};t)=\vec{\nabla}_{k}S(\{\vec{x}\};t)\label{eq:(26)}\end{equation}

\begin{equation}
P\times\vec{J}_{k}=Im[\Psi^{*}\vec{\nabla}_{k}\Psi]\label{eq:(27)}\end{equation}

\begin{equation}
P=R^{2}\label{eq:(28)}\end{equation}

\begin{equation}
Q(\{\vec{x}\};t)\equiv-\sum_{k}\frac{1}{2m_{k}}\frac{\nabla_{k}^{2}R(\{\vec{x}\};t)}{R(\{\vec{x}\};t)}\label{eq:(29)}\end{equation}

The last quantity introduced is known as the '' Quantum Potential''
or rather, {}``Quantum Information Potential'' (QIP) {[}2] and it
is, together with definition (\ref{eq:(26)}), the core of the alternative
formulation of non-relativistic quantum mechanics proposed by Bohm
and his co-workers {[}2,3,4]. As shown by these authors, it can be
naturally generalised to Quantum Field Theory (QFT).

In any case, we emphasize that all the above definitions are derived
from the time-dependent Schroedinger equation, and that their fundamental
status stands or falls with the validity of this equation.

Let us first introduce the other set of further relevant $beables$
{[}2,4] in BP, i.e. particles and fields: this $completes$ the formulation
of Quantum Theory in the BP.

One begins by defining dynamical quantities such as particle momenta,
in complete agreement - in the appropriate limit of course, i.e. when
Q is negligible - with the classical definition of momenta that is
part of the Hamilton-Jacobi formulation of classical mechanics {[}2,3].
We thus get Bohm's equations of motions for particles

\[
\vec{p}_{k}(\{\vec{X}(t)\};t)=m_{k}\frac{d\vec{X}_{k}(\{\vec{X}(t)\};t)}{dt}=\]

\begin{equation}
=\{\vec{\nabla}_{k}S(\{\vec{x}\};t)\}_{\{\vec{x}\}=\{\vec{X}(t)\}}\label{eq:(30)}\end{equation}

or (equivalently) the velocity of a particle k at its position $\vec{X}_{k}(t)$:

\[
\frac{d\vec{X}_{k}(\{\vec{X}(t)\};t)}{dt}=\]

\begin{equation}
=\frac{1}{m_{k}}[\frac{1}{P}Im(\Psi^{*}\vec{\nabla}_{k}\Psi)]_{\{\vec{x}\}=\{\vec{X}(t)\}}\label{eq:(31)}\end{equation}

This non-linear set of ordinary differential equations is the fundamental
set of Bohm equation of motion for $beable$ particles {[}2]: it is
the essential part of the theoretical framework that , according to
Bohm and co-workers, is missing from standard textbook formulations
and alternative ontological interpretations {[}2]. It is then demonstrated
that a complete removal of all the dead wood of ambiguities and {}``mysteries''
of textbook QM can thereby be elegantly achieved.

Eq.(\ref{eq:(28)}) can be rearranged to give an alternative version:

\[
m_{k}\frac{d^{2}\vec{X}_{k}(t)}{dt^{2}}=\]

\begin{equation}
=-[\vec{\nabla}_{k}(V(\{\vec{x}\})+Q(\{\vec{x}\};t))]_{\{\vec{x}\}=\{\vec{X}(t)\}}\label{eq:(32)}\end{equation}

One can thus clearly see that any beable particle is subject to $quantum$
$accelerations$, regardless of the existence of classical (inertial)
accelerations.

This sharply illuminates the profound difference between what a {}``particle''
means in classical physics and what a {}``beable particle'' means
in Bohm's quantum physics {[}2]. Unawereness of, or misunderstandings
about this fundamental difference can only lead to unnecessary confusions
and endless irrelevant objections and production of paradoxes.

The individual Bohm beable particle history (i.e. existing regardless
of whether the particle is, or is not, a member of an ensemble) is
then formally defined to be a solution of these non-linear ordinary
differential equations (31),(32).

Key features of quantum theory, such as quantum non-locality and quantum
wholeness {[}2], are thus precisely expressed by these equations for
particle (and field) motions as they are ultimately interpreted in
spacetime:

(i) single particle histories are primarily completely entangled (i.e.
$in$ $principle$ defined by, unseparable from ) the entire environment
(literally, the entire universe), because they are $\textit{actively }$$quantum$
$in-formed$ by the $phase$ of the wavefunction {[}2]. This has of
course nothing to do with $classically$ defined {}``medium entanglement''
(many-body effects) through {}``potentials, gravitation, thermal
influences'' and so forth; 

(ii) an essential feature of this quantum information is the de Broglie
$guidance$ condition on the particles eq (\ref{eq:(30)}), which
puts the $potential$ beable information encoded in the common wave
function into the $actual$ beable motion of every $individual$ beable
particle (as distinct from statistical ensembles of such particles),
participating in the potentially present common pool of information
provided by the wavefunction {[}2]. 

So, the physics at the quantum level of accuracy differs totally from
that assumed and expressed by classical mechanics, first and foremost
because it is this $quantum$ $wholeness$ that appears to be the
primary reality {[}2]. This is encoded $ab$$initio$ in the origin,
structure and every feature of the QIP {[}2].

It should be quite clear that one is indeed introducing here crucial
novel concepts that altogether single out the BP.

The beable Bohm histories of particles (and fields) then go $continuously$
over to $observable$ newtonian (or Hamilton-Jacobi) trajectories.
But one should be in no doubt {[}2] that the presence of the QIP profoundly
changes the entire situation from the bottom up, in a way that can
be neither understood nor interpreted in classical mechanics terms.
Thus the emergence of familiar classical spacetime descriptions is
indeed an immediate consequence of certain especial features of the
QIP. 

Two extreme situations can be immediately envisaged:

(i) The QIP is numerically completely negligible relative to the classical
potential (or, in more physical terms, compared to unavoidable thermal
fluctuations). In this case, the quantum forces derived from it are
completely overrun by ordinary classical mechanical and thermodynamical
forces.

A $\textit{continuous transition}$ to a classical regime (i.e. pure
spacetime descriptions, with use of coordinates) then occurs. Numerical
quantum corrections are then too small to play any important role.

(ii) The QIP is separable, i.e.

\[
Q(\{\vec{x}\};t)=\]

\begin{equation}
=-\sum_{k=1}^{N}\frac{1}{2m_{k}}\frac{1}{R_{k}(\{\vec{x}\};t)}\nabla_{k}^{2}R(\{\vec{x}\};t)\equiv\sum_{k=1}^{N}Q_{k}(\{\vec{x}\};t)\label{eq:(33)}\end{equation}

In this case, the specific quantum mechanical non-locality dissolves
and each particle behaves classically as if completly unaware of its
surroundings, except that it may interact with these surroundings
through eventual classically definable preassigned potentials, or
more physically speaking, through the omnipresent thermal fluctuations. 

POSTULATE 2:

In order to confront the BP (one should never forget that BP deals
primarily with individual beables {[}2,3,4]) with any existing experimental
material, one needs first to analyse in detail the anatomy of actual
measurement processes. 

This was thoroughly done by de Broglie, Bohm and their co-workers
{[}2,3,4] in general, at least for non-relativistic systems. The situation
when relativivistic requirements are essential is still far less digested,
both in textbook QFT and in the relativistic BP {[}2,3,4].

Let us then simply summarize one of the main results : one has to
apply some definite stochastic averaging procedures to the above theory
of individual beables, very much the same way as it is and was always
done in a classical context, e.g. in classical Statistical Mechanics
{[}2,3,4].

Given these individual Bohm histories of beables, it is straightforward
to compute statistical averages (i.e. averaging all initial positions
and velocities) over pure ensembles of such histories {[}2,3] . For
example, assuming for simplicity just one spatial dimension,

\begin{equation}
<X(t)>_{EA}\equiv\int dX(t)R^{2}(X(t),t)X(t)\label{eq:(34)}\end{equation}

\begin{equation}
<P(t)>_{EA}\equiv\int dX(t)R^{2}(X(t),t)[\frac{\partial S(x,t)}{\partial x}]_{x=X(t)}\label{eq:(35)}\end{equation}

\begin{equation}
<E(t)>_{EA}\equiv\int dX(t)R^{2}(X(t),t)[-\frac{\partial S(x,t)}{\partial t}]_{x=X(t)}\label{eq:(36)}\end{equation}

where the history $X(t)$ is given by (\ref{eq:(30)}). This turns
out to be - of course! - exactly the same as ordinary quantum mechanical
expectation values, taken with the $same$ wavefunction. 

So, statistics (a practical necessity, simply because $we$ cannot
know, nor is it necessary $for$ $us$ to know, everything that happened,
happens and shall happen in the world!) does not have in the context
of the present formulation the same fundamental status given to it
by textbook formulations of quantum theory, which deal solely with
{}``observables'', i.e. with the actual experimental data resulting
from actually performed experiments.

Most of these observables are context dependent, because of the participatory
nature of {}``observations'' in Quantum Mechanics {[}2]. Perhaps
the one important exception to this fact of life is the observable
{}``position of things'' {[}4], i.e the eigenvalues of the spatial
position operator. This observable can be directly linked to the beable
{}``position of (discrete) particles in space'' or {}``continuous
fields at specified points in space'', introduced in the previous
postulate. Space is, after all, where we exist and do experiments!
It is the playground of classical physics (and of common sense).

As both de Broglie and Bohm demonstrated, both mathematically and
using physics, the probability distributions to do this necessary
stochastic averaging, whatever they might originally be, rapidly relax
to none other than the familiar Born probabilities {[}2]. Because
of the local conservation of probability (\ref{eq:(25)}), if this
holds at some particular time, it will hold for all times. Hence,
within the context of any given real experiment, the results predicted
in principle by textbook and the ensemble averaged BP over all initial
positions $and$ velocities must be exactly the same.

A misunderstanding (or unawareness) of this most important POSTULATE
2 has caused a great deal of unnecessary confusion in the published
literature {[}5] about the meaning and status of the BP.

Real differences would be seen only for histories of individual systems,
as distinct from ensembles of such systems. Only if, and when, someone
some day has gathered sufficient such data can one make meaningful
judgements on the status of BP.

\section{Case studies from contemporary Nuclear and Particle physics}

We consider simple, but illustrative, additional examples inspired
by contemporary practices in Nuclear and Particle Physics. 

These fields of research offer excellent new oppurtunities for testing
specific interpretations and predictions based on the BP of Quantum
Theory. A good reason is, they often draw attention to the understanding
of two key features of all (quantum) processes, viz. the question
of transitions in time and the closely related feature of quantum
non-locality.

\subsection{EPR-like correlations in pure $\alpha-$decay channels of very light
nuclei}

Among the very light nuclei, the $^{8}Be$ and $^{12}C$ isotopes
can decay into pure two or three $\alpha-$channels with appreciable
branching ratios. These channels always played an important role in
the theories and experiments on ordinary stellar evolution. Thus the
momenta and energy distributions of the final state particles have
been thoroughly investigated {[}6,7,8], but not the predictable EPR-like
quantum correlations - the main issue here. The point of interest
is that the outgoing $\alpha's$ in these decays are bosons, even
if the parent nuclei are bags of nucleons, which are fermions. So
the wavefunctions in the outgoing decay channels must be fully symmetric
under exchanges of arbitrary labels attached to them (in this case,
just their position coordinates). So, any model candidate to a beable
wavefunction in the Center of Mass (CM) reference frame of this nucleus
must be a fully symmetric function of the internal coordinates.

We shall ignore for simplicity the final state Coulomb interaction
{[}7] that in principle operates among these outgoing particles and
accelerates them away from each other. We could assume, as it is done
in actual numerical fits {[}7], that this Coulomb interaction is screened;
in any case, it plays only a secondary role here, at it leads basically
to what could be called $classical$ $entanglement$.

However, the outgoing $\alpha-$channels, even if not interacting
through pre-assigned classical potentials, are EPR-correlated through
the Bose symmetry requirement on the wavefunction {[}2]. As is well-known
these EPR-correlations are essentially non-local {[}2], and thus beyond
any classical accountability.

We shall show from the viewpoint of the BP of Quantum Mechanics how
these correlations ( that must be there) can be traced back to the
symmetry requirement on the wavefunction and, as a by-product, attempt
to give a precise meaning to words such as {}``identical, equivalent,
distinguishable, undistinguishable,...'' particles.

In order to make this discussion as simple as possible but without
losing generality, let us consider only one spatial dimension and
imagine a $^{8}Be$ nucleus (at rest when t=$0$) fissioning into
two $\alpha$-particles. Extensions to heavier nuclei decays involving
more $\alpha-$ channels {[}6] do not add essentially new physics,
but do complicate the detailed mathematics considerably.

So, let the Hamiltonian for t>0 be

\begin{equation}
H=-\frac{1}{2M_{\alpha}}{\displaystyle \sum_{k=1}^{2}}\frac{\partial^{2}}{\partial x_{k}^{2}}\;\;\;\;\; t>0\label{eq:(37)}\end{equation}
 A central feature of this Hamiltonian is its invariance under the
two permutations of the labels 1,2 as a result of a common mass for
the particles. True stationary states must then share this symmetry
with the Hamiltonian. Quantum Mechanics establishes that these must
be fully symmetric functions of the coordinates, as one is dealing
with bosons {[}10]. The Schroedinger equation is

\begin{equation}
H\Psi(x_{1},x_{2};t)=i\frac{\partial}{\partial t}\Psi(x_{1},x_{2};t)\;\;\; t>0\label{eq:(38)}\end{equation}

It must be complemented with the initial condition

\begin{equation}
\Psi(x_{1},x_{2};0+)=F(x_{1},x_{2})\label{eq:(39)}\end{equation}

where F is a given (fully symmetric) function of spatial coordinates.
It is assumed of course that the wavefunction is continuous at t=0.

The consequences of the symmetry requirement are obvious, if we transform
to Jacobi coordinates:

\begin{equation}
x=x_{1}-x_{2}\label{eq:(40)}\end{equation}

\begin{equation}
X_{CM}=\frac{x_{1}+x_{2}}{2}\label{eq:(41)}\end{equation}

The coordinate $x$ is antisymmetric under the exchange $1\leftrightarrow2$
whereas $X_{CM}$ is symmetric. Hence we must have,

\begin{equation}
\Psi(x_{1},x_{2};t)=g(x;t)\times e^{iPX_{CM}}\label{eq:(42)}\end{equation}

where

\begin{equation}
g(x;t)=g(-x;t)\label{eq:(43)}\end{equation}

Let us assume that we are in the CM coordinate system First consider
a single particle guided by the wave-packet

\[
\Psi(x,t;\{A\})=\int_{-\infty}^{+\infty}exp(-i\frac{p^{2}}{2M_{\alpha}}t)*e^{-(p-p_{A})^{2}/\sigma_{A}^{2}}*\]

\begin{equation}
*exp(ip(x-x_{A})dp\;\;\;\; t>0\label{eq:(44)}\end{equation}

Putting

\begin{equation}
\Psi(x,t;\{A\})=R(x,t;\{A\})exp(iS(x,t;\{A\}))\label{eq:(45)}\end{equation}

we find that the particle velocity at time t is given by the guidance
condition (\ref{eq:(30)}):

\[
\frac{dX(t;\{A\}))}{dt}=\frac{1}{M_{\alpha}}\{\frac{\partial}{\partial x}S(x,t;\{A\}))\}_{x=X(t)}=\]

\begin{equation}
=\frac{1}{1+(\frac{\sigma_{A}^{2}}{2M_{\alpha}}t)^{2}}[\frac{p_{A}}{M_{\alpha}}+\frac{\sigma_{A}^{2}}{M_{\alpha}}(X(t;\{A\})-x_{A})t]\label{eq:(46)}\end{equation}

A simple quadrature then gives the particle position at time t. This
shows that the particle is accelerated from its initial velocity $p_{A}/M_{\alpha}$
at time t=0 to its final velocity at t=+$\infty$.

The fundamental reason for this pure quantum effect is, as discussed
in Section 1, the action of the QIP( \ref{eq:(29)} ) that is directly
linked to the Schroedinger wavefunction.

Next, consider two $\alpha's$ but ignore the symmetry requirement.
Then the wavefunction is simply

\begin{equation}
\Psi(x_{1},x_{2},t;\{A,B\})=\Psi(x_{1},t;\{A\})\times\Psi(x_{2},t;\{B\})\label{eq:(47)}\end{equation}

hence

\begin{equation}
R(x_{1},x_{2},t;\{A,B\})=R(x_{1},t;\{A\})\times R(x_{2},t;\{B\})\label{eq:(48)}\end{equation}

\begin{equation}
S(x_{1},x_{2},t;\{A,B\})=S(x_{1},t;\{A\})+S(x_{2},t;\{B\})\label{eq:(49)}\end{equation}

Using the guidance condition (\ref{eq:(30)}) the velocities are

\begin{equation}
\frac{d}{dt}X_{1}(t;A)=\{\frac{\partial}{\partial x_{1}}S(x_{1},t;\{A\})\}_{x_{1}=X_{1}(t)}\label{eq:(50)}\end{equation}

\begin{equation}
\frac{d}{dt}X_{2}(t;B)=\{\frac{\partial}{\partial x_{2}}S(x_{2},t;\{B\})\label{eq:(51)}\end{equation}

We conclude that the $\alpha-$particle that was labelled 1 at time
t=0 follows the beable trajectory A, i.e. $X_{1}(t;A)$, and is therefore
still the same particle at any later time t, exactly as one classically
would expect ; the same applies to particle 2, moving along the beable
spacetime trajectory $X_{2}(t;B)$. As expected classically, 1 and
2 are completely {}``unaware'' of each other, in spite of the fact
that they have a common origin, that is, the fissioning of $^{8}Be$.
There is no question about which is 1 and which is 2, as one can always
in principle follow their trajectories at any time. This applies,
even if 1 and 2 are {}``indistinguishable, equivalent, ...''. We
are so far in complete harmony with classical physics thinking.

Next, consider again two $\alpha's$ but insist that the symmetry
requirement must be obeyed by the wavefunction, as it should. Then
the correct wavefunction is

\[
\Psi(x_{1},x_{2},t;\{A,B\})=\frac{1}{\sqrt{2}}\{\Psi(x_{1},t;\{A\})\times\Psi(x_{2},t;\{B\})+\]

\[
+\Psi(x_{2},t;\{A\})\times\Psi(x_{1},t;\{B\})\}\]

\begin{equation}
=\Psi(x_{2},x_{1},t;\{A,B\})\label{eq:(52)}\end{equation}

Now the separability expressed by the definitions (\ref{eq:(48)})
and (\ref{eq:(49)}) is lost, and the results (\ref{eq:(50)}) and
(\ref{eq:(51)}) are no longer valid. Then the more general condition
(\ref{eq:(30)}) prevails:

\[
\frac{d}{dt}X_{1}(t;A,B)=\]

\begin{equation}
=\{\frac{\partial}{\partial x_{1}}S(x_{1},x_{2},t;\{A,B\})\}_{x_{1},x_{2}=X_{1}(t),X_{2}(t)}\label{eq:(53)}\end{equation}

\[
\frac{d}{dt}X_{2}(t;A,B)=\]

\begin{equation}
=\{\frac{\partial}{\partial x_{2}}S(x_{1},x_{2},t;\{A,B\})\}_{x_{1},x_{2}=X_{1}(t),X_{2}(t)}\label{eq:(54)}\end{equation}

This is tantamount to quantum non-locality, here meaning dependence
on $x=x_{1}-x_{2}$ that persists no matter how large $x$ is, even
in absence of any classically describable forces.

So, a natural (and classically meaningful) question would be: suppose
one places an $\alpha-$detector at a position D$_{A}$ and simultaneously
another identical $\alpha$-detector at a position D$_{B}$. These
detectors duly detect at some time t the two $\alpha-$particles upon
their arrival there. But is the particle that is detected at D$_{A}$
the {}``same'' particle called 1 at time t=0, and the particle detected
at D$_{B}$the {}``same'' called 2 at time t=0? The exact and real
answer is given by (\ref{eq:(53)}) and (\ref{eq:(54)}) which, within
the classical physics philosophy (and within the philosophy of the
usual common language as well), are meaningless!

Another feature of results (\ref{eq:(53)}) and (\ref{eq:(54)}) that
are closer to the original EPR-correlation problem, is: instead of
position detectors we could try momentum detectors. Eqs.(\ref{eq:(53)})and
(\ref{eq:(54)}) show that the $momentum$ of say particle 1 at time
t must depend on the $position$ at the $same\; time\; t$ of particle
2, independently of the distance separating the two particles and
vice-versa. Note that this conclusion applies to individual beables
and thus has nothing to do with the usual probabilities and statistics.
However, upon taking appropriate ensemble averages (Section 1), i.e.
doing the usual statistics of measurements {[}2,3,4], we would end
up exactly with the standard textbook predictions, which apply of
course only to some actual experimental result and refer only to that
particular experimental set up that produces that result.

\subsection{Neutron interferometry under gravitational fields}

The subject of specific quantum features seen in accelerated reference
systems is a promising field for future experimental research. Generally
speaking, ultracold neutron physics {[}9] is especially interesting,
due to high-quality experimental work now available.

Essentially quantum effects are deeply linked with changes in the
phase of the wavefunction {[}10]. So, the most revealing tell-tale
of the underlying quantum reality is most simply and directly provided
by interference experiments. Classical examples are provided by the
Aarhanov-Bohm and the Aarhanov-Casher effects {[}2,3].

More recently, very precise experimental results on motions of neutrons
in the earth's gravitational field became available {[}9]. Consider
a free neutron in motion under the gravitational field of the earth
{[}9,10]. Classically, any massive particle would move along the newtonian
trajectory given by a solution of the equation of motion

\begin{equation}
m_{in}\frac{d^{2}z}{dt^{2}}=-m_{gr}g\label{eq:(55)}\end{equation}

where the z-axis is chosen to be parallel to the gravitational lines
of force. As well-known {[}11] 

\begin{equation}
m_{in}/m_{gr}=universal\;\; constant\label{eq:(56)}\end{equation}

So a choice of mass units allows this universal constant to be set
to unity, and we get a simple $geometrical$ law of motion under gravity

\begin{equation}
\frac{d^{2}z(t)}{dt^{2}}=-g\label{eq:(57)}\end{equation}

\begin{equation}
z(t)=z(0)-\frac{1}{2}gt^{2}\label{eq:(58)}\end{equation}

quite independent of the mass parameter.

When using QM to reexamine this phenomenon, this most remarkable independence
on mass seems to disappear. However, this holds only if $\hbar\neq0$,
which of course is the case: the Schroedinger equation of motion is
then

\begin{equation}
[-\frac{d^{2}}{dz^{2}}-2(\frac{M_{N}}{\hbar})^{2}gz]\Psi(z,t)=2i(\frac{M_{N}}{\hbar})\frac{\partial}{\partial t}\Psi(z,t)\label{eq:(59)}\end{equation}

where $M_{N}$is the neutron mass. So the mass does contribute, but
only in the combination $M_{N}/\hbar$.

Note that there is no contradiction at all with the experimental fact
{[}11], just as there isn't any in classical mechanics.

Thus the Bohm equation of motion for any individual neutron that replaces
(\ref{eq:(57)}) can be written down. The neutron velocity is given
by eq

\begin{equation}
v(Z(t),t)=\frac{1}{M_{N}}[\frac{d}{dz}S(z,t)]_{z=Z(t)}\label{eq:(60)}\end{equation}

where $S(z,t)/\hbar$ is the phase of the wavefunction $\Psi(z,t)$.
From this the neutron history can be predicted, given the initial
conditions.

Consider now the following experiment {[}10]: a neutron beam with
energy E is split into two channels, a channel ABD and another channel
ACD. The channels are made to meet again at the neutron detector D.
For example, in the table-top experiment discussed in {[}10] the neutron
following the path ABD first climbs from the ground level up an inclined
plane (say, at angle $\delta$ with the ground) from A to B along
a path of length $l_{1}$then at B follows the path BD (length $l_{2}$)
parallel to the ground (thus normal to the gravitational lines of
force) until it reaches the detector at D. If a neutron is instead
guided along the path ACD, the opposite sequence is followed, i.e.
first AC (corresponding to BD) then CD (corresponding to AB) until
the neutron reaches the detector at D. The detectors at D record the
arrival at D of any neutron member of a statistical ensemble, sharing
the same wavefunction, as a function of the inclination $\delta$
for a given beam energy E.

This is of course just another example of a {}``which-path'' experiment,
this time using neutron interferometry {[}10]. Let us then assume
that

\[
\Psi(Z(t),t)=R_{1}(Z(t),t)exp(iS_{1}(Z(t),t)/\hbar)+\]

\begin{equation}
+R_{2}(Z(t),t)exp(iS_{2}(Z(t),t)/\hbar)\label{eq:(61)}\end{equation}

The probability pattern at the detector is then

\begin{equation}
P(Z(t),t)=R_{1}^{2}(Z(t),t)+R_{2}^{2}(Z(t),t)+2R_{1}R_{2}cos\Phi_{12}\label{eq:(62)}\end{equation}

where $\Phi_{12}$ is the phase difference between the two paths : 

\begin{equation}
\Phi_{12}=(S_{1}(ABD)-S_{2}(ACD))/\hbar\label{eq:(63)}\end{equation}

It can be quickly estimated in the WKB approximation (which, remember,
neglects the QIP). Phase change along ABD:

\begin{equation}
\frac{S_{ABD}}{\hbar}=\int_{ABD}pdl/\hbar=\int_{ABD}\sqrt{2M_{N}(E-M_{N}gz)}dl/\hbar\label{eq:(64)}\end{equation}

and along ACD:

\begin{equation}
\frac{S_{ACD}}{\hbar}=\int_{ACD}pdl/\hbar=\int_{ACD}\sqrt{2M_{N}(E-M_{N}gz)}dl/\hbar\label{eq:(65)}\end{equation}

hence

\begin{equation}
\Phi_{12}=-l_{2}\sqrt{2M_{N}E}/(2E\hbar)*M_{N}gl_{1}sin\delta\label{eq:(66)}\end{equation}

Recalling that

\begin{equation}
\sqrt{2M_{N}E}/E=2M_{N}\lambda/h\label{eq:(67)}\end{equation}

where $\lambda$ is the de Broglie wavelength of the incoming wavefunction.
We thus get our final result {[}10]

\begin{equation}
\Phi_{12}=-(M_{N}/\hbar)^{2}gl_{1}l_{2}sin\delta*\frac{\lambda}{2\pi}\label{eq:(68)}\end{equation}

This prediction appears to be in excellent agreement with the experimental
results {[}9,10].

Let us now consider the realistic case that any incoming neutron with
given velocity belongs in a statistical ensemble of such particles. 

Upon arrival at time t=0 at A they are all redirected (i.e. the trajectories
bifurcate). Depending on the initial position and velocity of a particle,
either the path ABD or the path ACD will be followed, but every single
body is forced to meet at D. 

The histories Z(t) of individual neutrons are given by formula (30).The
actually observed pattern of arrivals of individuals at D is fixed
by (62).

The $probability$ that at time t=0 any particular neutron $is$ at
Z=0 with velocity V(0) is $assumed$ (Postulate 2) to be given by 

\begin{equation}
P(Z(0)=0,t=0)=|\Psi(Z(0)=0,t=0)|^{2}\label{eq:(69)}\end{equation}

We emphasize that in BP there is nothing fundamental about this assumption.

As a matter of fact, this is the only place where probabilities enter
the game at all, just as they do in Classical Statistical Mechanics.
Everything else for t>0 is completely determined by the initial conditions.
Thus the probability that a neutron $will$ $be$ at the position
Z(t) (with velocity V(t) given by (30 )) is

\begin{equation}
P(Z(t),t)=|\Psi(Z(t),t)|^{2}\label{eq:(70)}\end{equation}

in full formal agreement with textbooks.

\subsection{Quantum transitions in nuclear $\beta$-decay processes}

The ongoing experimental and theoretical work in Nuclear and Particle
Physics has by now initiated an exciting new era of fundamental physics.
Random samples of the enormous output of publications in later years
that are relevant to this paper are {[}9],{[}12],{[}13 -17],{[}18]).

An especially interesting and relevant feature of recent developments
is the strong light they cast on the nature of quantum transitions
in electroweak and QCD physics.

\subsubsection{The time-dependent wavefunction}

Good examples of contemporary interest is K-capture in highly ionized
H-like atoms {[}18] and the physics of ultracold neutrons in traps
{[}9]. We shall $reinterpret$ this according to the philosophy sketched
in this paper. Let us consider the decays

\begin{equation}
_{Z}A_{N}+e^{-}\rightarrow_{Z-1}A_{N+1}+\nu_{e}\label{eq:(71)}\end{equation}

This is a K-capture process, meaning capture of an 1S atomic electron
by a proton in a parent nucleus $_{Z}A_{N}$ with Z protons and N
neutrons through electroweak interactions that instantaneously replace
a proton with a neutron in the daughter nucleus $_{Z-1}A_{N+1}$ ,
together with an outgoing electron neutrino $\nu_{e}$. This is accompanied
by a sudden, but in this case rather small, change of the Coulomb
field of the nucleus. An interesting feature of this kind of experiments
is that the parent atom is supposed to be H-like, i.e. to be completely
ionized except for a single bound electron, say in the 1S atomic state
{[}18].

We shall consider this without any explicit or implicit reference
to {}``meaurements'', {}``probabilities'' , ''observations''
or any other related standard concepts and terminology used in textbooks
{[}2,3,4], where they are considered to be essential for self-consistency.
As already discussed, this is not the case with the BP {[}2,3,4],
where {}``measurements'' of {}``observables'' (implying actually
no more than a permanent record of events stored somewhere and referring
exclusively to a specific experimental set-up) are just special cases
of quantum transition processes for beables {[}2,3,4]. Postulate 2
would link this to any specific available or predictable experimental
results and, of course, to any textbook postdictions or predictions
of such results. 

Again, this is not the purpose of the present paper.

We imagine a movie showing the time development of a K-capture process
ocurring say at time t=0.

The Hamiltonian driving the evolution of the wave function is assumed
to be

\begin{equation}
H=U_{0}+H_{0}+V_{eff}\label{eq:(72)}\end{equation}

$U_{0}$ represents the (infinite) Dirac vacuum energy. The next term
is defined as

\[
H_{0}=H_{nuc}-\frac{\hbar^{2}}{2m_{e}}\nabla_{y}^{2}(e)-Z\alpha\int d^{3}\vec{x}\frac{\varrho_{el}(\vec{x})}{|\vec{x}-\vec{y}|}+\]

\begin{equation}
{\displaystyle +\sum_{n=1}^{3}}[-i\hbar\vec{\alpha}(n).\vec{\nabla}_{y}(n)+\beta(n)m_{n}c^{2}]\label{eq:(73)}\end{equation}
We shall reserve the notation $\{\vec{x}\}$ for space coordinates
of the nucleons. The symbol $\vec{y}$ refers exclusively to the spatial
coordinates of leptons. The first term on the rhs of (\ref{eq:(72)})
represents the nuclear Hamiltonian:

\[
H_{nuc}\Phi_{\alpha}(\{\vec{x},\vec{x}_{CM}\};Z,N)=\]

\begin{equation}
=E_{\alpha}(Z,N)\Phi_{\alpha}(\{\vec{x},\vec{x}_{CM}\};Z,N)\label{eq:(74)}\end{equation}

where 

\begin{equation}
\vec{x}_{CM}\equiv\frac{{\displaystyle {\displaystyle M_{P}\sum_{i=1}^{Z}\vec{x}_{iP}+M_{N}\sum_{i=1}^{N}\vec{x}_{iN}}}}{ZM_{P}+NM_{N}}\label{eq:(75)}\end{equation}

are the CM coordinates of the nucleus defined in the atomic CM reference
system.The next two terms represent the Hamiltonian of the single
atomic 1S electron with $\varrho_{el}(\vec{x})$ as the electric charge
density profile of the nucleus. In the non-relativistic approximation

\[
[-\frac{\hbar^{2}}{2m_{e}}\nabla_{y}^{2}(e)-Z\alpha\int d^{3}\vec{x}\frac{\varrho_{el}(\vec{x})}{|\vec{x}-\vec{y}|}]\varphi_{1S\sigma}(\vec{y})=\]

\begin{equation}
=-|\varepsilon_{1S}|\varphi_{1S\sigma}(\vec{y})\label{eq:(76)}\end{equation}

The last term in definition (73) stands for the Dirac Hamiltonian
of massive neutrinos, with the Bjorken-Drell conventions and definitions
{[}20]. We need actually only the positive energy solutions $u_{n\vec{p}\sigma}(\vec{y})$:

\[
[-i\hbar\vec{\alpha}(n).\vec{\nabla}_{y}(n)+\beta(n)m_{n}c^{2}]u_{n\vec{p}\sigma}(\vec{y})=\]

\begin{equation}
=\sqrt{|\vec{p}|^{2}c^{2}+m_{n}^{2}c^{4}}u_{n\vec{p}\sigma}(\vec{y})\label{eq:(77)}\end{equation}

The term $V_{eff}$ in definition (72) is the electroweak interaction
as originally suggested by the Standard Model {[}19] but adapted here
(as an effective interaction) to the physical conditions inside nuclei
{[}21]. As such it must be used with careful insight and according
to standard ideas on modern renormalization techniques and effective
field theories {[}21].

Assume that at time $t\leq0$ the initial wavefunction of an individual
H-like ion at rest in its CM system is

\[
\Psi_{I,1S\sigma}(\{\vec{x},\vec{x}_{M}\},\vec{y};t)\equiv<\{\vec{x},\vec{x}_{M}\},\vec{y}|I,1S\sigma;t>=\]

\begin{equation}
=\Phi_{I}(\{\vec{x},\vec{x}_{M}\};Z,N)\varphi_{1S\sigma}(\vec{y})e^{-iE_{I1S}t}\;\;\;\; t\leq0\label{eq:(78)}\end{equation}

where $\Phi_{I}(\{\vec{x},\vec{x}_{M}\};Z,N)$ is the internal wavefunction
of the parent nucleus. The total available energy is $E_{I1S}$:

\begin{equation}
E_{I1S}=U_{0}+{\displaystyle (}M_{I}(Z,N)+m_{e})c^{2}-|\varepsilon(1S)|\label{eq:(79)}\end{equation}

where M$_{I}$(Z,N) is the mass of the parent nucleus.

However, this state is not really a stationary state, because of the
background weak couplings of quarks inside the nucleons to virtual
flavour changing heavy electroweak bosons $W^{\pm},Z^{0}$ {[}21].
This is represented by the (hermitean) last term $V_{eff}$ on the
rhs of the effective Hamiltonian (\ref{eq:(72)}) for allowed transitions
{[}21]:

\begin{equation}
V_{eff}(\vec{x},\vec{y})=-\frac{G_{F}}{\sqrt{2}}\delta(\vec{x}-\vec{y})(V_{F}(n)+V_{GT}(n))\label{eq:(80)}\end{equation}

where $G_{F}=8.7\times10^{-5}MeV.fm^{2}$ is the Fermi constant ,
$V_{F}(V_{GT})$ are related to the standard Fermi (Gamov-Teller)
interaction operators. 

Let us make the following ansatz for the time-dependent quantum state
that should be good enough for sufficiently $weak$ $interactions$
{[}10]:

\[
|I,1S\sigma;t>=\]
\begin{equation}
=exp(-i(E_{I1S}-i\frac{\Gamma_{I1S}}{2})t)|I,1S\sigma;t=0>\;\;\;\Gamma_{I1S}>0\label{eq:(81)}\end{equation}

It can be easily shown that {[}10]

\[
\Gamma_{I1S}=2\pi{\displaystyle \sum_{F}{\displaystyle \sum_{n,\sigma_{n}}\int d^{3}\vec{p}_{n}}}*\delta(E_{I1S}-E_{nF}(p_{n}))*\]

\begin{equation}
*|<F(-\vec{p}_{n}),n\vec{p}_{n}\sigma_{n}|V_{eff}|I,1S\sigma>|^{2}\label{eq:(82)}\end{equation}
in agreement with Fermi's Golden Rule {[}10]. Furthermore

\[
E_{nF}(p_{n})=\]

\[
=U_{0}+{\displaystyle (}M_{F}(Z-1,N+1)c^{2}+\frac{|\vec{p}_{n}|^{2}}{2M_{F}}+\]

\begin{equation}
+\sqrt{|\vec{p}|_{n}^{2}c^{2}+m_{n}^{2}c^{4}}\label{eq:(83)}\end{equation}

in obvious notation. The leptonic coordinates in the atomic CM system
are related to $\vec{x}_{CM}$ by the definition

\begin{equation}
m_{e}\vec{y}+M_{I}\vec{x}_{CM}=0\label{eq:(84)}\end{equation}

in the initial state and

\begin{equation}
m_{n}\vec{y}+M_{F}\vec{x}_{CM}=0\label{eq:(85)}\end{equation}

in the final state.

We have used momentum conservation when writing down the expression
for the recoil energy of the final nucleus.

The Bohm survival probability for an individual parent at any time
t>0 is defined to be

\begin{equation}
P_{I1S}(t)=<I,1S\sigma;t|I,1S\sigma;t>=exp(-\Gamma_{I1S}t)P_{I1S}(0)\label{eq:(86)}\end{equation}

This must satisfy the probability conservation law (25). So it does,
to first order, since $V_{eff}$ is by construction hermitean. Let
us define as usual

\[
\Psi_{I1S\sigma}(\{\vec{x},\vec{x}_{CM}\},\vec{y};t)=R_{I1S\sigma}(\{\vec{x},\vec{x}_{CM}\},\vec{y};t)*\]

\begin{equation}
=*exp(iS_{I1S\sigma}(\{\vec{x},\vec{x}_{CM}\},\vec{y};t)\label{eq:(87)}\end{equation}

where

\[
R_{I1S\sigma}(\{\vec{x},\vec{x}_{CM}\},\vec{y};t)=\]

\begin{equation}
=exp(-\frac{\Gamma_{I1S}}{2}t)|\Psi_{I,1S\sigma}(\{\vec{x},\vec{x}_{CM}\},\vec{y},0)|\label{eq:(88)}\end{equation}
\begin{equation}
P_{I1S}(t)=R_{I1S\sigma}^{2}(\{\vec{x},\vec{x}_{CM}\},\vec{y};t)\label{eq:(89)}\end{equation}

Any built-in degeneracies of $\Psi_{I1S\sigma}(\{\vec{x},\vec{x}_{CM}\},\vec{y};0)$
would imply that its phase may not be zero.

\subsubsection{Particle histories}

In the context of the ontological BP under discussion a good question
(though one that is never asked by any textbook!) could be: what are
precisely the beable particle histories, i.e. the positions $and$
velocities of those inside our decaying initial ion and the outgoing
neutrino, at any time after t=0, until eventually some entirely new
conditions - such as an observation! - occur {[}2]?

In the context of the BP of Quantum Theory, not only the question
does make perfect sense, but can also be easily answered in quantitative
detail, at least in perturbation theory as sketched above (even if,
for the present, we are quite unable to plan, let alone carry out,
any experiments based on such ideas !).

Especially interesting is any possibly {}``measurable quantity'',
e.g. the momentum of the outgoing beable neutrino n or rather, the
recoil energy of the daughter nucleus. The precise answer given by
the BP is (section 1):

\[
\vec{p}_{n}(\{\vec{X}(t),\vec{X}_{CM}(t)\},\vec{Y}(t);t)=\]

\begin{equation}
=\{\vec{\nabla}_{\vec{y}}S_{I1S\sigma}(\{\vec{x},\vec{x}_{CM}(t)\},\vec{y};t)\}_{\{\vec{x}\}\vec{x}_{CM},\vec{y}=\{\vec{X}(t)\},\vec{X}_{CM}(t),\vec{Y}(t)}\label{eq:(90)}\end{equation}
 Immediate confrontation with existing experimental material on this
question is however possible ( Postulate 2 ,section 1) if one first
carries out an appropriate stochastic averaging over all initial positions
and velocities of the (entangled) beable nucleons and leptons initially
involved :

\[
<\vec{p}_{n}(\{\vec{X}(t),\vec{X}_{CM}(t)\},\vec{Y}(t);t)>_{EA}=\int\int\int[R_{I1S\sigma}^{2}(\{\vec{x}\}\vec{x}_{CM},\vec{y};t))*\]

\begin{equation}
*\vec{\nabla}_{y}(n)S_{I1S\sigma}(\{\vec{x}\}\vec{x}_{CM},\vec{y};t)\}_{\{\vec{x}\}\vec{x}_{CM},\vec{y}=\{\vec{X}(t)\}\vec{X}_{CM}(t),\vec{Y}(t)}\label{eq:(91)}\end{equation}

The integrations are to be carried out over all histories of all participating
beable particles. As mentioned in section 2, this is guaranteed to
be in numerical agreement with the textbook answer at t=0. As also
mentioned in section 2, it will then be true at any time :

\[
<(-i\vec{\nabla}_{y}(n)>=\]

\[
=\frac{1}{2}\int\int[\Psi_{I1S\sigma}^{*}(-i\vec{\nabla}_{y}(n)\Psi_{I1S\sigma})+\Psi_{I1S\sigma}(-i\vec{\nabla}_{y}(n)\Psi_{I1S\sigma})^{*}]\equiv\]

\begin{equation}
\equiv<\vec{p}_{n}(\{\vec{X}(t)\}\vec{X}_{CM}(t),\vec{Y}(t);t)>_{EA}\label{eq:(92)}\end{equation}

We emphasize that the word {}``time'' means here literally {}``time
taken by a particle, say the neutrino n, to be guided from its position
$\vec{X}(t)$ (and velocity $\vec{V}(t)$) to position $\vec{X}'(t')$
(and velocity $\vec{V}'(t')$)''. If a {}``distance'' L is {}``measured''
from the neutrino position to the CM of the parent atom , then at
time t>0 the distance L can be traded with time t elapsed because
to sufficient accuracy $t=L/c.$ Hence, the time label t in (\ref{eq:(91)})
can be replaced by the space label L, and so all talk about a possibly
vague {}``time'' can be avoided {[}22]. The point being, {}``time''
here is directly linked to the physical motions of material particles.

\section{Summary and conclusions}

The ontological formulation of non-relativistic Quantum Mechanics,
originally initiated by Louis de Broglie in the twenties but independently
recreated and reformulated in the fifties and sixties by D. Bohm and
his co-workers, is briefly presented and its explanatory powers are
further illustrated with examples from new fields of research in contemporary
Nuclear and Particle Physics, fields that were not mature during the
lifetime of those pioneers.

It is argued that the reformulation in question has indeed the capability
of not only by-passing all the so-called paradoxes and ambiguities
characterizing the official (textbook) epistemological formulations
of Quantum Theory, but more importantly, lead to genuine testable
$physical\;\; explanations$ (and not only to highly abstract mathematical
representations) of actual physical phenomena, at the quantum level
of accuracy. This is achieved without any need of arbitrary breaks
of continuity between the so-called {}``classical'' and {}``quantal''
modes of description that is so vital for all textbook formulations.
Furthermore, this novel ontological formulation can lead naturally
to proposals for new type of experimentation, thus in principle greatly
extending the predictive powers of present day Quantum Theory. One
can thus open up for new ways of thinking and trying fresh approaches
to the age-old fundamental problems of Relativistic Quantum Field
Theory and Theoretical Cosmology.

\end{document}